\documentclass[12pt,dvips]{article}
\usepackage{epsfig}
\title{\bf 
Low energy scattering and photoproduction of $\eta$-mesons on deuterons. }

\author{N. V. Shevchenko, V. B. Belyaev \\
{\it  Joint Institute  for Nuclear Research, Dubna, 141980, Russia} \\
\\
  S. A. Rakityansky, S. A. Sofianos \\
{\it  Physics Dept., University of South Africa, P.O. Box 392 } \\
 {\it  Pretoria 0003, South Africa} \\
\\
       and W. Sandhas \\
{\it Physikalisches Institut, Universit\"{a}t Bonn, D-53115 Bonn, 
Germany}}
\date{}
\begin{document}
\maketitle

\section{Introduction}
The production of $\eta$ mesons and their collisions with nuclei have
been studied experimentally and theoretically with increasing interest
during the last years. To a large extent this is motivated by the 
fundamental problem of charge--symmetry breaking of strong 
interaction~\cite{Magi00}. Another relevant 
questions concerns the nature of $S_{11} (1535)$--resonance~\cite{Hoeh98} 
and the possible
formation of $\eta$--nucleus quasi--bound states~\cite{Hai86}. 
It is worth mentioning that according 
to~\cite{Rol96} the mean free path of $\eta$ mesons in a
nuclear medium is about 2\,fm, {\em i.e.}, less than the size of a 
typical nucleus. A necessary condition for the existence of 
$\eta$-nuclei, hence, appears to be satisfied.

For the  calculation of these states various model treatments were 
employed, among them the optical potential 
method~\cite{ref4,ref6}, the Green's function method~\cite{ref7},
the modified multiple scattering theory~\cite{gnw} and few--body
calculations~\cite{Ueda} -- \cite{She98}.
The predictions concerning the possibility of $\eta$--mesic nucleus 
formation are very diverse. One obvious reason for such a diversity 
is the poor knowledge of the $\eta N$ forces. Another reason comes 
from the differences among the employed approximations some of which 
might be faulty in view of the resonant character of the 
$\eta N$ dynamics and the delicacy of the quasi--bound state problem.

In the present paper we treat the $\eta$--deuteron system on the
basis of the exact few--body Alt--Grassberger--Sandhas (AGS) 
equations. The Faddeev--type coupling of these equations guarantees 
uniqueness of their solutions.  Moreover, as equations for the 
elastic and rearrangement operators they are well--defined in 
momentum space, providing thus the desired scattering amplitudes 
in a most direct and technically reliable manner.  
The advantage of working with coupled equations involving the elastic
and rearrangement operators of the final state, is not only
suggested by questions of uniqueness, but also by the relevance of
rescattering effects which were found to give a significant
contribution to the corresponding amplitude~\cite{Rit99}.

\section{AGS formalism}
In terms of the AGS transition operator $U_{11}$ the $\eta d$
elastic scattering amplitude is represented as
\begin{equation}
\label{ampl}
	  f({\bf p}_1',{\bf p}_1;z)=-(2\pi)^2 M_1
	  \langle {\bf p}_1';\psi_d|U_{11}(z)|{\bf p}_1;\psi_d\rangle
\end{equation}
with the on--energy--shell conditions  $|{\bf p}_1'| = |{\bf p}_1|$
and $z = p_1^2/2 M_1 + E_d$ with $E_d$ being the deuteron energy.
Here the subscript 1 labels the $\eta(NN)$ partition and the 
$\eta$--deuteron channel.
The transition operator $U_{11}$ obeys the system of AGS equations
\begin{equation}
\label{ags}
U_{\beta \alpha}(z) = (1 - \delta_{\beta \alpha}) G_{0}^{-1}(z) +
\sum_{\gamma=1}^3 (1-\delta_{\beta \gamma}) T_{\gamma}(z) G_{0}(z)
U_{\gamma \alpha}(z),
\end{equation}
with $G_{0}(z)$ being the free resolvent (Green's operator) of the
three particles involved. This set of equations couples all $3
\times 3$ elastic and rearrangement operators $U_{\alpha \alpha}$
and $U_{\beta \alpha}$. Here each of the subscripts runs through
the values 1, 2 and 3, indicating the two-fragment partitions
(1,23), (2,31) and (3,12) respectively. Therefore, $U_{11}$
describes the elastic transition $1(23)\to 1(23)$, while $U_{21}$
represents the rearrangement process $1(23)\to 2(13)$. These
subscripts are also used in the complementary notation to label
the two--body T--operator $t_{\alpha}(z)$ of the ($\beta \gamma$)
pair, for instance $t_1(z) = t_{NN}(z)$. It should be noticed,
however, that it is not this genuine two--body operator which
enters the AGS equations, but the operator
\begin{equation}
\label{equ4}
T_{\alpha}(z) = t_{\alpha}(z - {\bf q}_{\alpha}^2/2 M_{\alpha}),
\end{equation}
which is to be understood as the two--body operator embedded in
the three--body space, with the relative kinetic energy operator
${\bf q}_{\alpha}^2/2 M_{\alpha}$ of particle $\alpha$ being
subtracted from the total energy variable z. Considered in
momentum space,~(\ref{equ4}) thus reads
\begin{equation}
\langle {\bf p'}_{\alpha}, {\bf q'}_{\alpha}| T_{\alpha}(z)| {\bf
p}_{\alpha}, {\bf q}_{\alpha} \rangle = \delta({\bf q'}_{\alpha}
-{\bf q}_{\alpha}) \langle {\bf p'}_{\alpha}| t_{\alpha}(z -
q_{\alpha}^2/2 M_{\alpha})|
 {\bf p}_{\alpha} \rangle,
\end{equation}
where ${\bf p}_{\alpha}$ and ${\bf q}_{\alpha}$ are the Jacobi
momenta of the pair $\alpha$ and the spectator $\alpha$
respectively, and $M_{\alpha}$ is the corresponding reduced mass.

For both $T_{\eta N}$ and $T_{NN}$ we used one--term separable forms
\begin{equation}
\label{sept}
       T_{\alpha}(z)=|\chi_{\alpha}\rangle \tau_{\alpha}(z) 
\langle\chi_{\alpha}|\ .
\end{equation}
For the $NN$ subsystem Eq.~(\ref{sept}) implies that the asymptotic wave
function is related to the form-factor $|\chi_1\rangle$ according to
\begin{equation}
\label{dfun}
  |{\bf p}_1;\psi_d\rangle = G_0(z)|\chi_1\rangle|{\bf p}_1\rangle\ .
\end{equation}
Due  to~(\ref{sept}) and~(\ref{dfun}) the scattering 
amplitude~(\ref{ampl}) can be rewritten as
\begin{equation}
\label{ampl1}
	  f({\bf p}_1',{\bf p}_1;z) = -(2\pi)^2 M_1
	  \langle {\bf p}_1'|X_{11}(z)|{\bf p}_1\rangle\ ,
\end{equation}
where the operators $X_{\beta \alpha}$, defined as
$$
 X_{\beta \alpha}(z)=\langle\chi_{\beta}|G_0(z) U_{\beta \alpha}(z) G_0(z)
|\chi_{\alpha}\rangle\ ,
$$
obey the system of equations
\begin{equation}
\label{agsx}
     X_{\beta \alpha}(z) = Z_{\beta \alpha}(z) + 
\sum^3_{\gamma = 1} Z_{\beta \gamma}(z) \, \tau_{\gamma} 
\left(z - \frac{p_{\gamma}^2}{2 M_{\gamma}}\right) \, X_{\gamma \alpha}(z)
\end{equation}
with
$$
  Z_{\beta \alpha}(z)=(1 - \delta_{\beta \alpha})
\langle\chi_{\beta}|G_0(z)|\chi_{\alpha}\rangle\ .
$$
The identity of the nucleons implies that $X_{31}=X_{21}$, $\tau_3=\tau_2$,
and $Z_{31}=Z_{21}$, which reduces the system~(\ref{agsx}) to two coupled
equations
\begin{equation}
\label{agsi}
\begin{array}{rcl}
  X_{11}(z) & = &
  \displaystyle
  2 Z_{12}(z)\tau_2\left(z-\frac{p_2^2}{2 M_2}\right) X_{21}(z)\ , \\
  &\\
  X_{21}(z) & = & Z_{21}(z)+
  \displaystyle
  Z_{21}(z) \tau_1\left(z-\frac{p_1^2}{2 M_1}\right)X_{11}(z)+
  Z_{23}(z) \tau_2\left(z-\frac{p_2^2}{2 M_2}\right)X_{21}(z)\ .\\
\end{array}
\end{equation}
Eventually, after making the $S$--wave projection of the matrix 
elements
$\langle{\bf p}_{\beta}'|X_{\beta \alpha}|{\bf p}_{\alpha}\rangle$ and
$\langle{\bf p}_{\beta}'|Z_{\beta \alpha}|{\bf p}_{\alpha}\rangle$, 
we end up with
one--dimensional integral equations which can be solved numerically.

\section{Two--body T--matrices}
The $S$-wave nucleon-nucleon separable potential is adopted from
Ref.\cite{garcilazo} with its parameters slightly modified  to be 
consistent with more recent $NN$ data (see~Ref.\cite{She98}). 
The $\eta$-nucleon $T$-matrix is taken in the form
\begin{equation}
\label{tetan}
     t_{\eta N}(p',p;z)=({p'}^2+\alpha^2)^{-1}
    \frac{\lambda}{(z-E_0+i\Gamma/2)}(p^2+\alpha^2)^{-1}
\end{equation}
consisting of two vertex functions and the $S_{11}$-propagator in 
between~\cite{Rak96}.  It corresponds to the process 
$\eta N\to S_{11}\to \eta N$
which at low energies is dominant. The range parameter $\alpha=3.316$
fm$^{-1}$ was determined in Ref.~\cite{bennh}, while $E_0$ and $\Gamma$ 
are the parameters of the $S_{11}$ resonance~\cite{PDG},
$$
E_0=1535\,{\rm MeV}-(m_N+m_\eta)\ ,\qquad \Gamma=150\,{\rm MeV}\ .
$$
The strength parameter $\lambda$ is chosen to reproduce the
$\eta$-nucleon scattering length $a_{\eta N}$,
\begin{equation}
\label{t000}
  \lambda=\frac{\alpha^4(E_0-i\Gamma/2)}{(2\pi)^2\mu_{\eta N}}a_{\eta N}\ .
\end{equation}
the imaginary part of which accounts for the flux losses into 
the $\pi N$ channel.

The two--body scattering length $a_{\eta N}$
is not accurately known. Different
analyses \cite{batinic} provided for $a_{\eta N}$ the values in the range
\begin{equation}
\label{interval}
0.27\ {\rm fm}\le{\rm Re\,}a_{\eta N}\le 0.98\ {\rm fm}\ ,\qquad
0.19\ {\rm fm}\le{\rm Im\,}a_{\eta N}\le 0.37\ {\rm fm}\ .
\end{equation}
Recently, however, most of the authors agreed that Im $a_{\eta N}$ 
is around $0.3$ fm.  But for Re $a_{\eta N}$ the estimates are still 
very different (compare, for example, Refs.~\cite{finn} and~\cite{speth}).
Most of the results presented in this paper are, therefore, obtained 
using Im $a_{\eta N} = 0.3$ fm and
several values of Re $a_{\eta N}$ within the above interval.

We solved Eqs.~(\ref{agsi}) for  $\eta d$ collision energies varying 
from zero ($\eta d$--threshold, $z = E_d$) up to $22$ MeV by
replacing the integrals by Gaussian sums. As is well known (see, for
example,~\cite{Bel}), the kernels of these equations, when expressed in
momentum representation, have moving logarithmic singularities for 
$z > 0$. In the numerical procedure, we handle it with the method 
suggested in Ref.\cite{Soh71}. The main idea of this method consists 
in interpolating the unknown solutions (in the area covering the 
singular points) by certain polynomials and subsequent analytic 
integration of the singular part of the kernels.

\section{Scattering results}
\begin{figure}[h]
\begin{minipage}[t]{80mm}
\includegraphics[width=70mm,height=50mm]{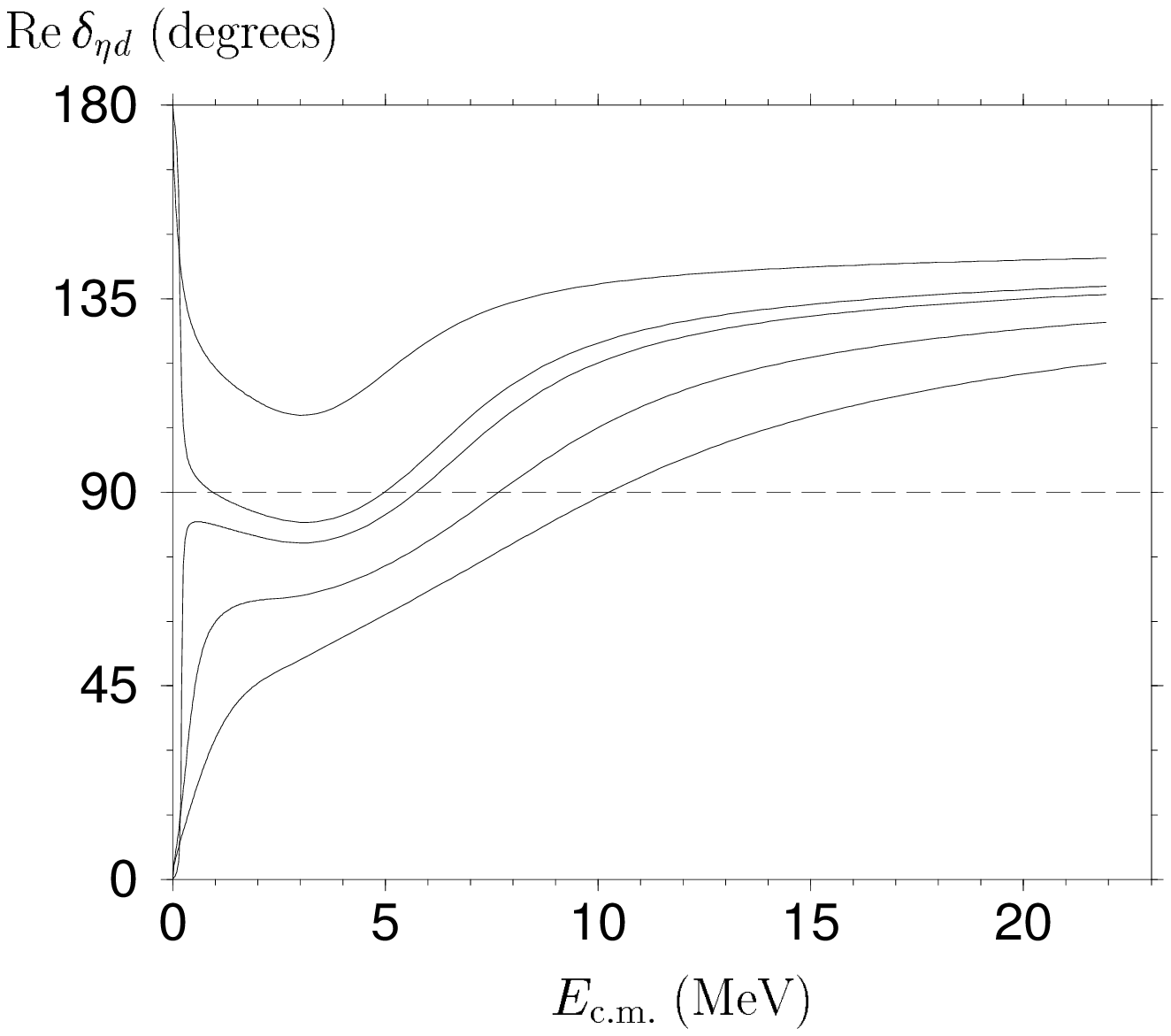}
\caption{Real part of the $\eta$-deuteron phase--shift as a function 
of the collision energy.
}
\label{red.fig}
\end{minipage}
\hspace{\fill}
\begin{minipage}[t]{75mm}
\includegraphics[width=70mm,height=50mm]{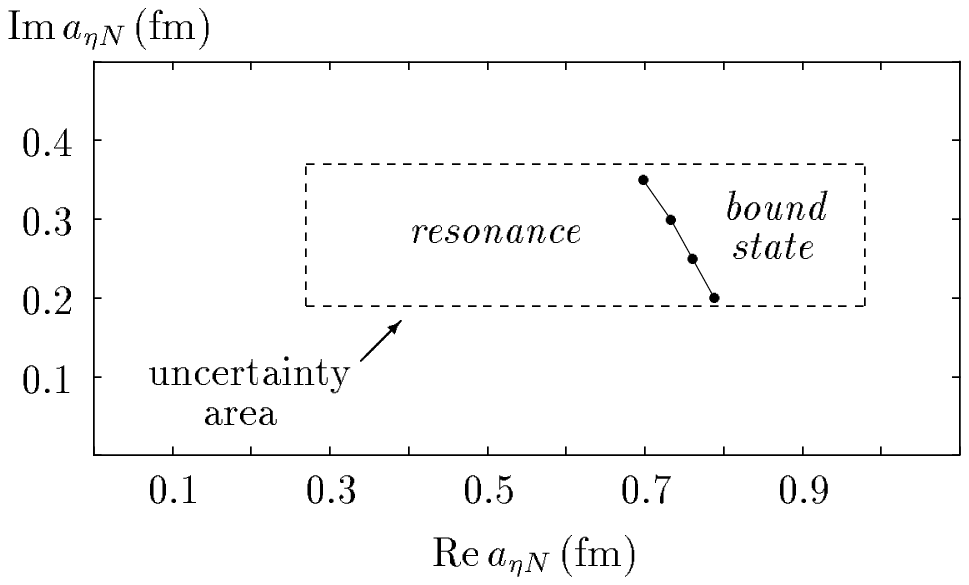}
\caption{The critical  values of $a_{\eta N}$ (filled circles) to the right 
of which the $\eta d$--system can be bound.
}
\label{area.fig}
\end{minipage}
\end{figure}

\begin{figure}[h]
\begin{minipage}[t]{80mm}
\includegraphics[width=70mm,height=50mm]{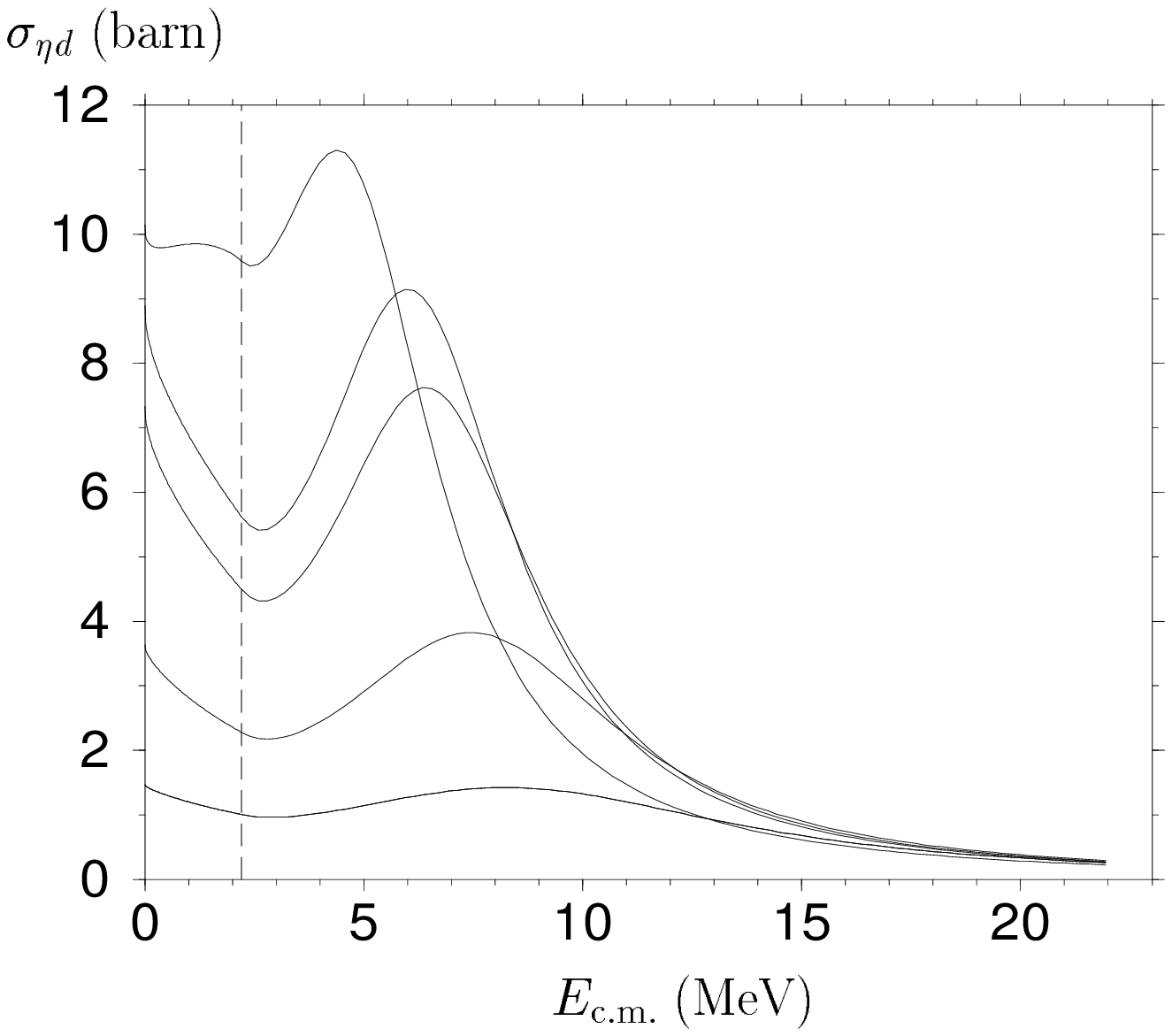}
\caption{Total cross--section for elastic $\eta d$ scattering
as a function of collision energy.
}
\label{cross.fig}
\end{minipage}
\hspace{\fill}
\begin{minipage}[t]{75mm}
\includegraphics[width=70mm,height=50mm]{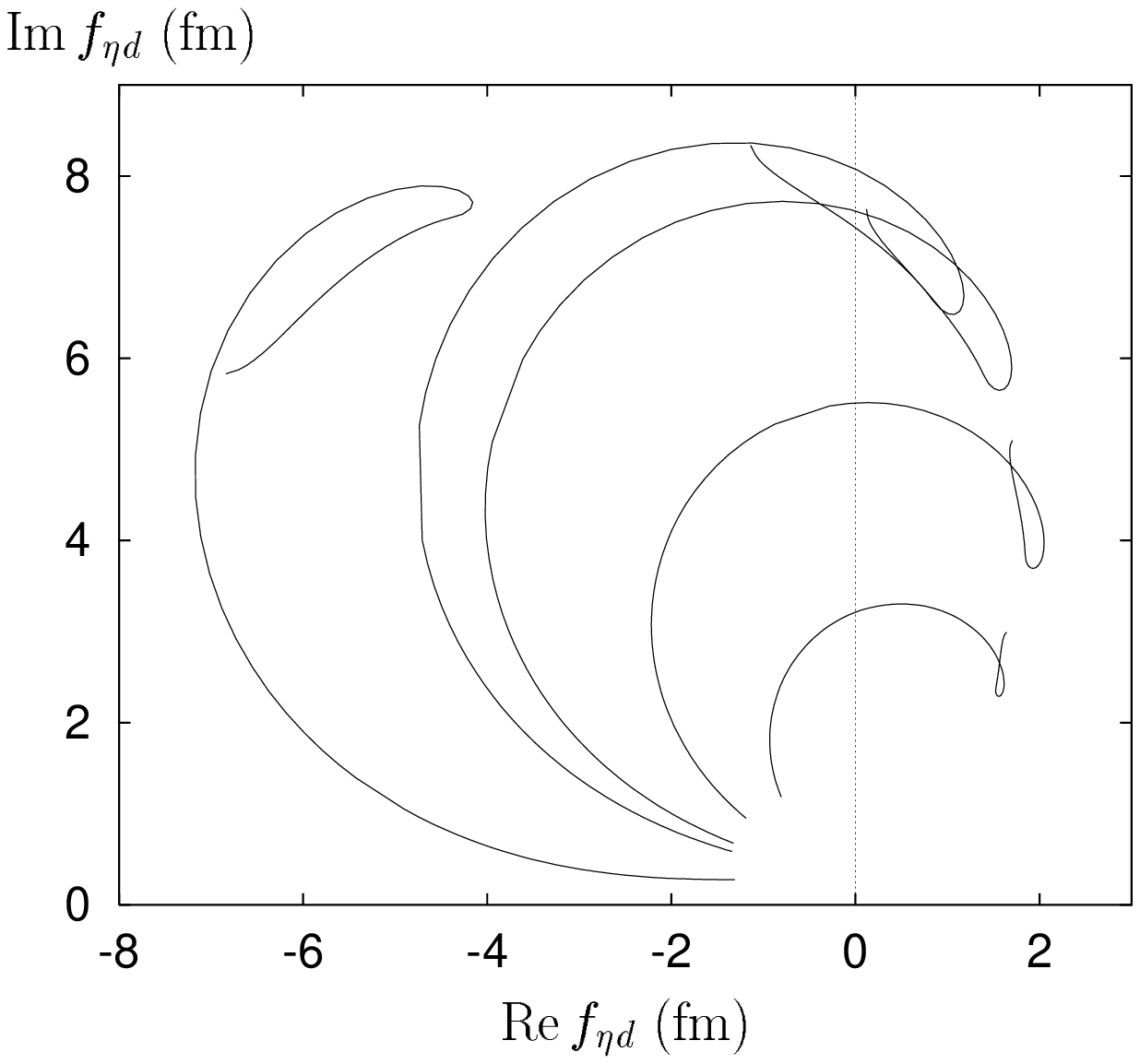}
\caption{Argand plot for the $\eta d$ elastic scattering amplitude.
 }
\label{argand.fig}
\end{minipage}
\end{figure}

The results of our calculations are presented in
Figs.~\ref{red.fig}--\ref{argand.fig} and in Table~\ref{res.tab}. In
Fig.~\ref{red.fig} the energy dependence of the $\eta d$
phase-shifts for five different choices of Re $a_{\eta N}$ is shown.
The curves correspond (starting from the lowest one) to 
Re $a_{\eta N} = 0.55$ fm, $0.65$ fm, $0.725$ fm, $0.75$ fm, and $0.85$ fm.
The larger this value, the stronger is the $\eta$N attraction.  
The change in the character of these curves, hence, reflects the 
growth of the attractive force between the $\eta$ meson and the 
nucleon.  The lower three curves for Re $\delta_{\eta d}$ corresponding 
to the smaller values of Re $a_{\eta N}$ start from zero, the two 
curves corresponding to the strong attraction start from $\pi$.  
According to Levinson's theorem, the phase shift at threshold energy 
is equal to the number of bound states $n$ times $\pi$.  We found 
that the transition from the lower family of the curves to the upper 
one happens at the critical value Re $a_{\eta N} = 0.733$ fm. Therefore, 
the $\eta$N force, which generates Im $a_{\eta N}=0.3$ fm and
Re $a_{\eta N} > 0.733$ fm, is sufficiently attractive to bind $\eta$ 
inside the deuteron.  

For three other choices of Im $a_{\eta N}$ within the uncertainty 
interval, namely, $0.20$ fm, $0.25$ fm, and $0.35$ fm, the corresponding
critical values of Re $a_{\eta N}$ turned out to be $0.788$ fm, 
$0.761$ fm, and $0.698$ fm.  
In the complex $a_{\eta N}$--plane (see Fig.~\ref{area.fig}) the 
corresponding points form a curve separating the uncertainty area 
(given by formulae~(\ref{interval}), dashed rectangle)
into two parts. If $a_{\eta N}$ is to the right of this curve, 
the strength of $\eta$N attraction is sufficient for $\eta d$ bound 
state formation.
\begin{table}[h]
\begin{center}
\begin{tabular}{|c|c|c|}
\hline
 Re $a_{\eta N}$ (fm) &    $E_{\eta d}^{\rm res}$ (MeV) &
 $\Gamma_{\eta d}$ (MeV)\\
\hline
   0.55          &             8.24        &        9.15  \\
   0.65          &             7.46        &        8.45  \\
   0.675         &             7.14        &        7.61  \\
   0.70          &             6.79        &        6.90  \\
   0.725         &             6.41        &        6.31  \\
   0.75          &             6.01        &        5.87  \\
   0.85          &             4.39        &        5.79  \\
   0.90          &             3.73        &        6.81  \\
\hline
\end{tabular}                    
\caption{Energy and width of the $\eta d$ resonance for various 
choices of Re $a_{\eta N}$.}
\label{res.tab}
\end{center}
\end{table}

In Fig.~\ref{cross.fig} we present the result of our calculations
of the total cross-section  (integrated over the angles) for elastic
$\eta$-deuteron scattering as a function of collision energy. The five
curves correspond (starting from the lowest one) to 
Re $a_{\eta N} = 0.55$ fm, $0.65$ fm, $0.725$ fm, $0.75$ fm, and $0.85$ fm. 
The dashed line indicates the deuteron break--up threshold.
The peaks in the energy dependence of the total elastic 
cross--section indicate that a resonance appears in the 
$\eta d$--system. Of course, not every
maximum of the cross--section is a resonance, but 
we plotted the Argand plots also~(Fig.~\ref{argand.fig})
for the $\eta d$ elastic scattering amplitude in the
energy interval from $0$ to $22$ MeV. The five curves correspond 
(from right to left) to Re $a_{\eta N} = 0.55$ fm, $0.65$ fm, $0.725$ fm, 
$0.75$ fm, and $0.85$ fm. When the energy increases the corresponding 
points move anticlockwise. So that the Argand plots prove that the 
maxima we found are resonances. 
Their positions and widths for various choices of 
Re $a_{\eta N}$ are given in Table~\ref{res.tab}. It should be noted 
that, while the resonance energy is determined in our calculations 
exactly (as the maximum of the function $\sin^2{\rm Re\,}\delta_{\eta d}$),
 the corresponding width is obtained by fitting the cross-section 
with a Breit-Wigner curve.  Therefore, the values of $\Gamma_{\eta d}$ 
given in Table~\ref{res.tab} should be considered only
as rough estimates.

\medskip
The resonant behavior of $\eta d$ elastic scattering should be seen 
in various processes involving $\eta d$-system in their final states, 
such as $\gamma d \to \eta d$ and $ n p \to \eta d$.  Indeed, the 
corresponding amplitudes $\langle\psi_{\rm out}|{\cal O}|\psi_{\rm in}\rangle$ 
involve the $\eta d$ wave function $\psi_{\rm out}$ which, in the 
vicinity of the resonance, strongly depends on the total energy. 
Its resonant growth at short distances may enhance the transition 
probability. To check this suggestion we performed the calculations 
of the $\gamma d \to \eta d$ reaction.

\section{Photoproduction of $\eta$-mesons}
Theoretical analysis of $(\gamma,\eta)$-reactions on nuclei is hampered 
by the three major problems: the unknown off-shell behavior of the
two-body $\gamma N \to \eta N$ amplitude, inaccuracies in the
description of the nuclear target as a many-body system, and
rescattering effects in the final state.
The simplest is, of course, the process of coherent
$\eta$ photoproduction on deuteron. There are many theoretical
studies devoted to ($\gamma, \eta$) reactions on deuteron. Early
attempts to go beyond a simple impulse approximation 
led to very different conclusions~\cite{bib1} -- \cite{bib3} as do more recent
approaches based on the effective two-body
formulations~\cite{bib4},~\cite{bib5}. Moreover, the experimental
cross-section~\cite{bib6} of the reaction
\begin{equation}
\label{equn}
\gamma + d \to \eta + d
\end{equation}
in the near-threshold region is far above these theoretical
predictions. Therefore, a reliable description of $\eta$
photoproduction on deuteron 
on the basis of exact equations is desirable.

\section{Formalism}
To consider reaction~(\ref{equn}), we employ the exact
Alt-Grassberger-Sandhas formalism modified to include the
electromagnetic interaction.
 A photon can be introduced into this formalism by considering the
$\eta N$ and $\gamma N$ states as two different channels of the
same system. This means that we should replace the T-operator
$t_{\eta N}$  by the $2 \times 2$ matrix. It is clear, that such
replacements of the kernels of integral equations~(\ref{ags}) lead 
to the corresponding solutions having a similar matrix form
\begin{equation}
\label{tmatr} 
t_{\eta N} \to \left(
\begin{array}{cc}
 t^{\gamma \gamma} & t^{\gamma \eta} \\
 t^{\eta \gamma}   & t^{\eta \eta}
\end{array}
\right) \qquad
U_{\alpha \beta} \to \left(
\begin{array}{cc}
 W_{\alpha \beta}^{\gamma \gamma} & W_{\alpha \beta}^{\gamma \eta} \\
 W_{\alpha \beta}^{\eta \gamma}   & W_{\alpha \beta}^{\eta \eta}
\end{array}
\right).
\end{equation}
Here $t^{\gamma \gamma}$ describes the Compton scattering,
$t^{\eta \gamma}$ the photoproduction process, and $t^{\eta \eta}$
the elastic $\eta N$ scattering\footnote{Another method of treatment
electromagnetic process of the type~(\ref{equn}) in the frame of AGS 
equations was used in Ref.\cite{Fix}.}.

It is technically more convenient to consider the reaction of
$\eta$-photoabsorption, which is inverse to reaction~(\ref{equn}). Then 
the photoproduction cross-section can be obtained by applying the
detailed balance principle.
We are interested in the coherent process, therefore we need the 
amplitude $W_{11}^{\gamma \eta}$ obeying the equation
\begin{eqnarray}
\nonumber \left(
\begin{array}{cc}
 W_{11}^{\gamma \gamma} & W_{11}^{\gamma \eta} \\
 W_{11}^{\eta \gamma}   & W_{11}^{\eta \eta}
\end{array}
\right)
                                 &=&
\left(
\begin{array}{cc}
 T_2^{\gamma \gamma} & T_2^{\gamma \eta} \\
 T_2^{\eta \gamma} & T_2^{\eta \eta}
\end{array}
\right)
                              G_0
\left(
\begin{array}{cc}
 W_{21}^{\gamma \gamma} & W_{21}^{\gamma \eta} \\
 W_{21}^{\eta \gamma}   & W_{21}^{\eta \eta}
\end{array}
\right)                      \\ \label{system}
                   &{}& \;\; +
\left(
\begin{array}{cc}
 T_3^{\gamma \gamma} & T_3^{\gamma \eta} \\
 T_3^{\eta \gamma} & T_3^{\eta \eta}
\end{array}
\right)
                              G_0
\left(
\begin{array}{cc}
 W_{31}^{\gamma \gamma} & W_{31}^{\gamma \eta} \\
 W_{31}^{\eta \gamma}   & W_{31}^{\eta \eta}
\end{array}
\right).
\end{eqnarray}
In the first order on electromagnetic interaction:
\begin{equation}
\label{solve}
 W_{11}^{\gamma \eta} \approx T_2^{\gamma \eta} G_0(z)
W_{21}^{\eta \eta} +
 T_3^{\gamma \eta} G_0(z) W_{31}^{\eta \eta}
\end{equation}
and we can see, that in above approximation the transition operator
$W_{11}^{\gamma \eta}$ can be found not from an equation, but from
the expression~(\ref{solve}). It is physically clear since in the first
order on electromagnetic interaction the only contributions are from
the $\eta$--meson rescattering which is described by the
$W_{21}^{\eta \eta}$ and $W_{31}^{\eta \eta}$ operators.

It was experimentally proven~\cite{bib14} that at low energies the
reaction $\gamma N \to \eta N$ mainly goes via formation of the
$S_{11}$-resonance, which means that $t^{\gamma \eta}$ in the
near--threshold region can be written in a separable form similar
to~(\ref{sept}). To construct such separable T--matrix, we used the 
results of Ref.~\cite{bib15} where $t^{\gamma \eta}$ was obtained as an
element of a multi--channel T--matrix which simultaneously describes
data for the processes
\begin{eqnarray}
\nonumber
 &{}& \pi + N \to \pi + N,    \quad   \pi + N \to \eta + N, \\
\nonumber
 &{}& \gamma + N \to \pi + N, \quad  \gamma + N \to \eta + N
\end{eqnarray}
on the energy shell in the $S_{11}$--channel. For our calculations, we 
extended this T--matrix off the energy shell,
\begin{equation}
t_{off}^{\gamma \eta}(p',p;E) =
\frac{\kappa^2 + E^2}{\kappa^2 + {p'}^2} \; t_{on}^{\gamma \eta}(E) \;
\frac{\alpha^2 + 2 \mu E}{\alpha^2 + p^2},
\end{equation}
using the Yamaguchi form--factors which become unit on the energy
shell. Here $\kappa$ stands for some unknown parameter.
It is known, that $t^{\gamma \eta}$ is
different for neutron and proton, we assumed that they have the
same functional form and differ by a constant factor
$ t_n^{\gamma \eta} = A \, t_p^{\gamma \eta}.$
A multipole analysis~\cite{bib16} gives for this factor the 
following estimate: $ A = -0.84 \pm 0.15.$

\section{Photoproduction results}
\begin{figure}[h]
\begin{minipage}[t]{80mm}
\includegraphics[width=70mm,height=50mm]{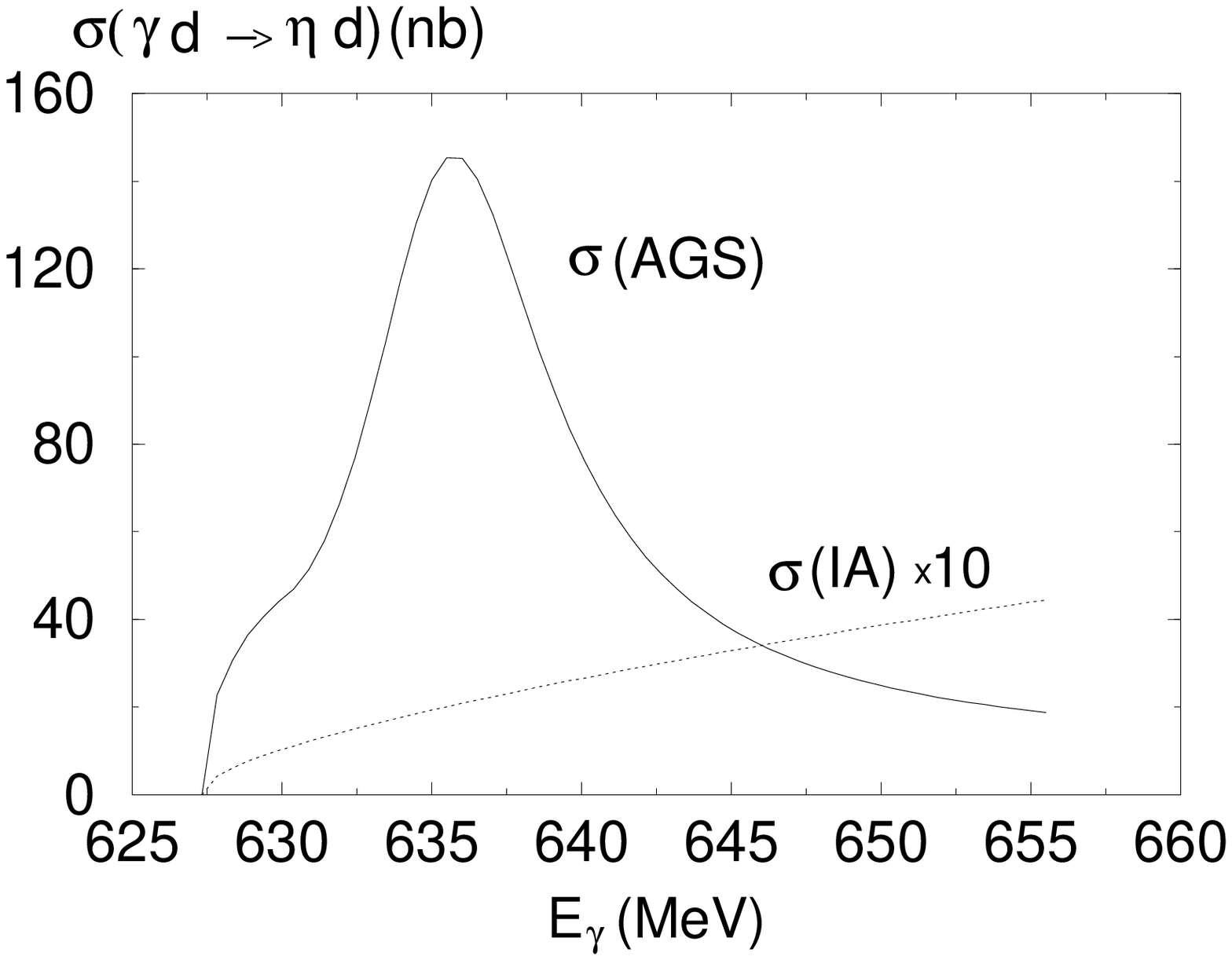}
\caption{
Total cross-section, calculated within a
rigorous few-body theory (AGS) and Impulse Approximation (IA).}
\label{imp.fig}
\end{minipage}
\hspace{\fill}
\begin{minipage}[t]{75mm}
\includegraphics[width=70mm,height=50mm]{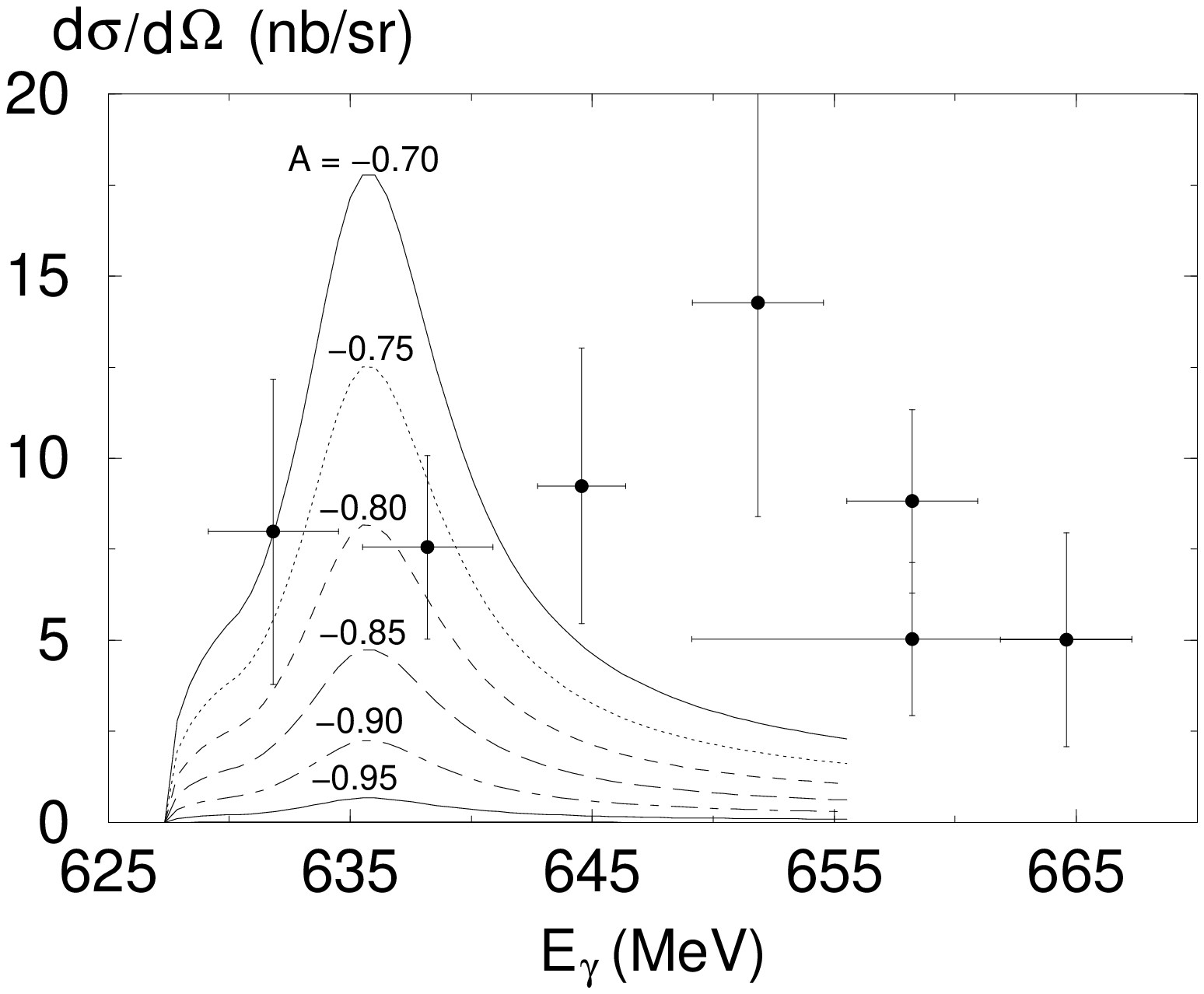}
\caption{
Differential cross-section ($\Theta_{\eta}^{cm} = 90^0$),
calculated with different choices of the ratio $A$. }
\label{fnfp.fig}
\end{minipage}
\end{figure}

\begin{figure}[htb]
\begin{minipage}[t]{80mm}
\includegraphics[width=70mm,height=50mm]{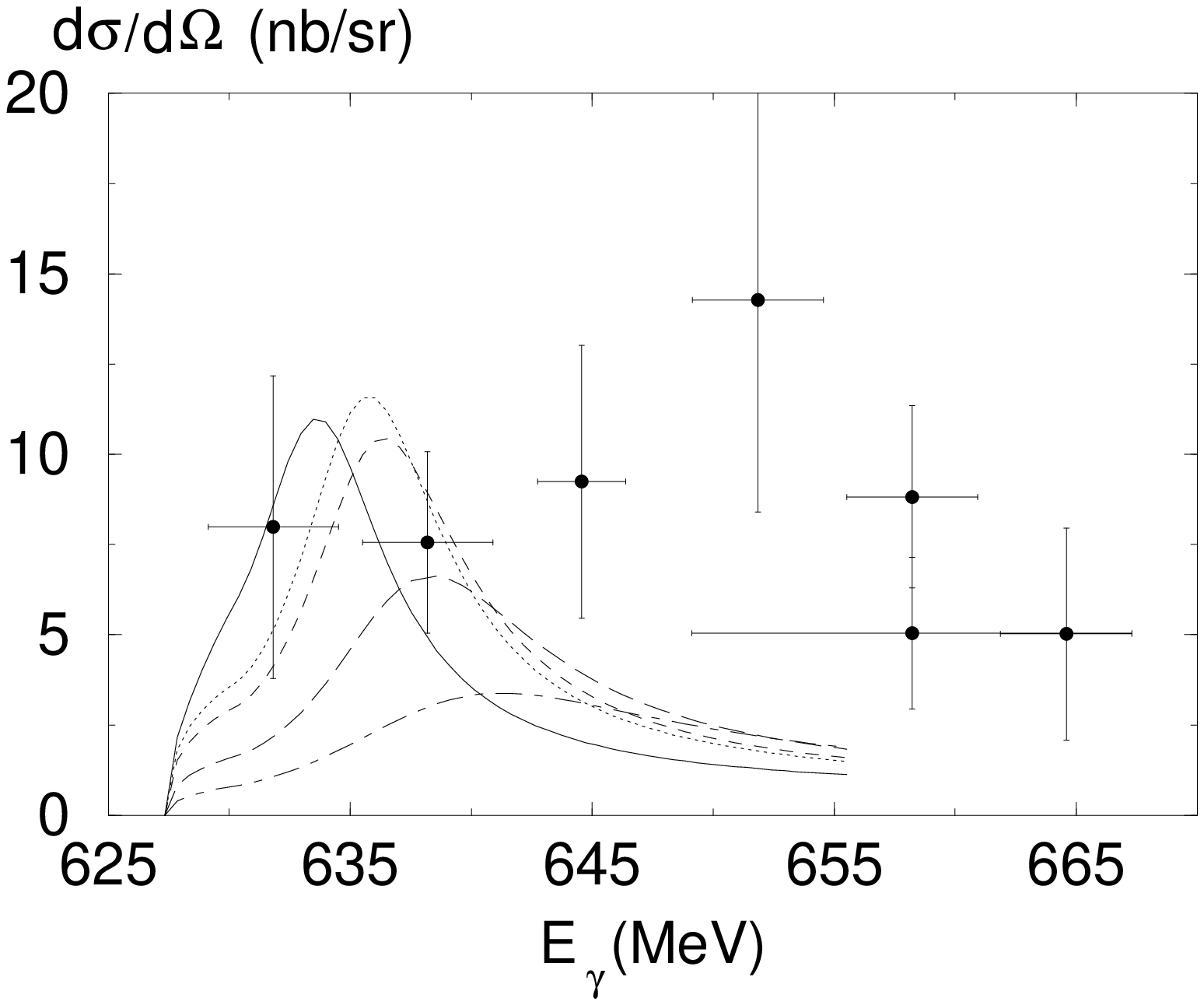}
\caption{
Differential cross-section ($\Theta_{\eta}^{cm} = 90^0$)
with different choices of Re $a_{\eta N}$. }
\label{rea.fig}
\end{minipage}
\hspace{\fill}
\begin{minipage}[t]{75mm}
\includegraphics[width=70mm,height=50mm]{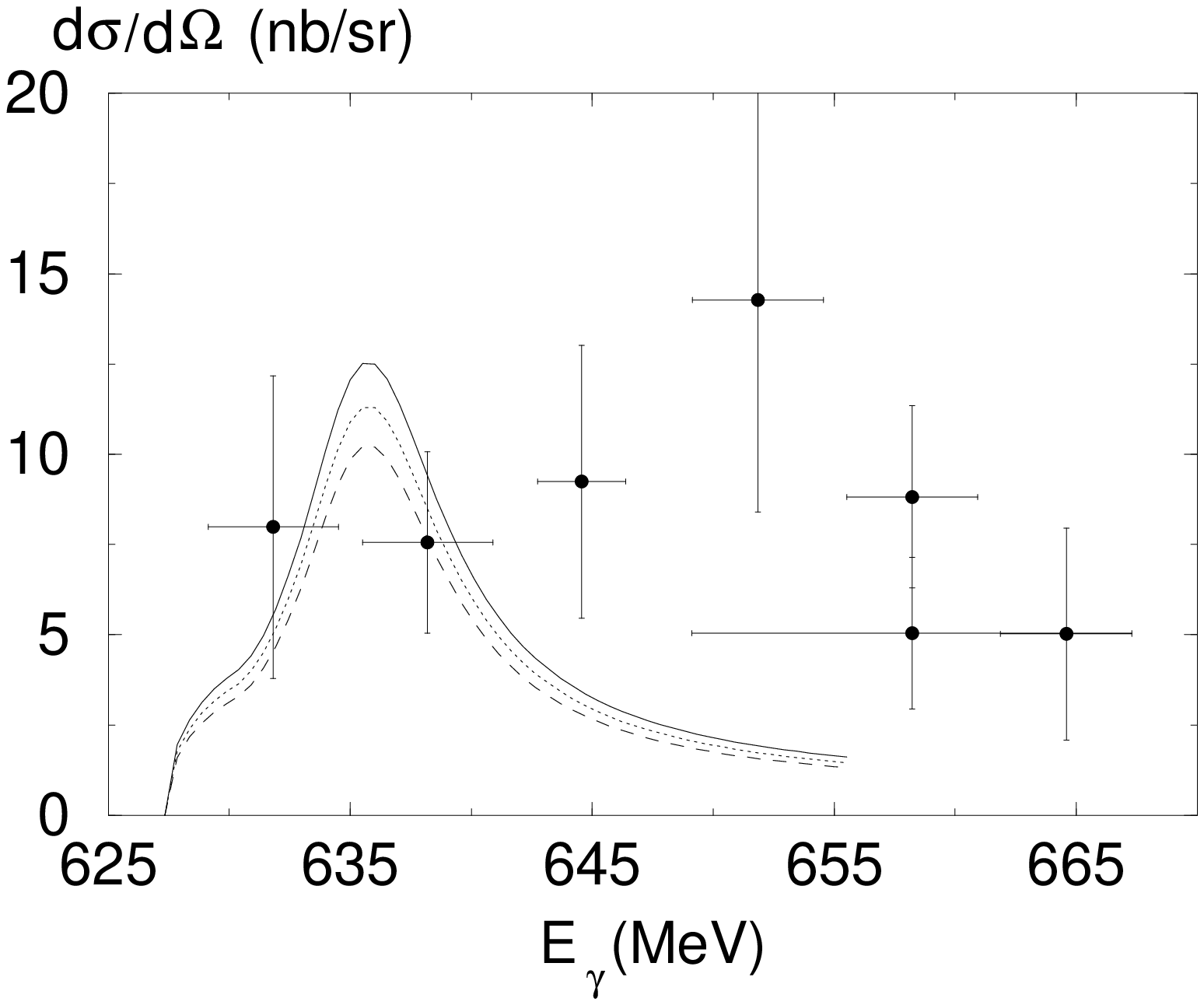}
\caption{
Differential cross-section 
with different choices of parameter $\kappa$.}
\label{kapa.fig}
\end{minipage}
\end{figure}

As was expected, our calculations revealed very strong final state
interaction in the reaction~(\ref{equn}). A comparison of the 
corresponding cross--sections obtained by solving the AGS equations 
and by using the Impulse Approximation (IA) is given in 
Fig.~\ref{imp.fig}, where the IA-results are multiplied by 10.
Besides the fact that the IA-curve is generally an order of
magnitude lower, it does not show a resonant enhancement which is
clearly seen when all the rescattering and re-arrangement
processes are taken into account. In this connection, it should be
noted that experimental data~\cite{bib6} for the reaction~(\ref{equn}),
given in Figs.~\ref{fnfp.fig} -- \ref{kapa.fig} show a pronounced 
enhancement of the differential cross--section at low energies.

In order to examine a dependence of our calculations on the choice
of the parameters of the T--matrices $t^{\eta \eta}$ and 
$t^{\gamma \eta}$, we did variations of 
$A = t_n^{\gamma \eta}/t_p^{\gamma \eta}$,  Re $a_{\eta N}$ and $\kappa$
within the corresponding uncertainty intervals. One of the most 
important parameters of the theory is the ratio of the 
photoproduction amplitudes for neutron and proton ($A$). Six curves 
corresponding to different choices of $A$ are depicted in 
Fig.~\ref{fnfp.fig}. These curves were calculated with 
$a_{\eta N} = (0.75, 0.30)$ fm and $\kappa = \alpha = 3.316$ fm$^{-1}$.
The experimental data are taken from Ref.~\cite{bib6}.

In Fig.~\ref{rea.fig}, we present the result of our calculations for 
five different choices of Re $a_{\eta N}$, namely, $0.55$ fm, $0.65$ fm,
$0.725$ fm, $0.75$ fm, and $0.85$ fm. This sequence of Re $a_{\eta N}$ 
corresponds to the upward 
sequence of the curves in the near--threshold region.
The parameters are: $A = -0.75$ and $\kappa = \alpha = 3.316$ fm$^{-1}$. 

In Fig.~\ref{kapa.fig} the results of our calculations with three 
choices of the range parameter $\kappa$ (2~fm$^{-1}$, 3~fm$^{-1}$, and 
5~fm$^{-1}$) of the electromagnetic vertex $\gamma N \to S_{11}$ are
given. These curves correspond to $a_{\eta N} = (0.75, 0.30)$ fm and
$A = -0.75$.
A comparison of the curves depicted in Figs.~\ref{fnfp.fig},~\ref{rea.fig}
and~\ref{kapa.fig} with the corresponding experimental data shows that 
no agreement with the data can be reached unless the ratio $A$ is 
greater than $-0.80$.
\begin{figure}[htb]
\begin{center}
\begin{minipage}[t]{80mm}
\includegraphics[width=70mm,height=50mm]{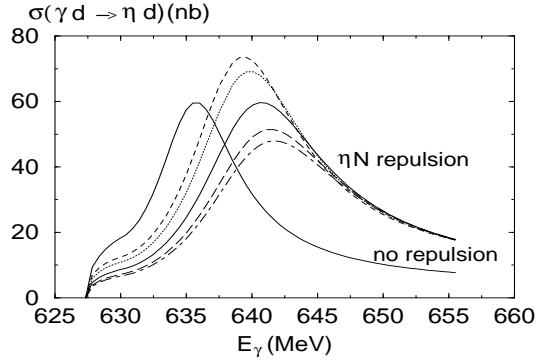}
\caption{
Shift of the resonant peak of the total cross-section 
due to the $\eta N$ repulsion.}
\label{repuls.fig}
\end{minipage}
\end{center}
\end{figure}

Under all variations of the parameters, however, the resonant peak
remains about 15 MeV to the left of the experimental peak. Since
this peak is due to the resonant final state interaction between
the $\eta$ meson and deuteron, we may expect that it can be
shifted to the right by introducing a repulsion into the $\eta N$
interaction. To introduce an $\eta N$ repulsion which preserves the 
separable form of the corresponding T--matrix, we used the method 
suggested in Ref.~\cite{garcilazo} where a separable nucleon--nucleon 
T--matrix includes an energy dependent factor,
$ b(E) = -\tanh{( 1 - E/E_c )}. $
This factor causes the NN phase--shift to change sign at the energy
$E_c = 0.816$ fm$^{-1}$, which is equivalent to presence of an NN
repulsion. Since the purpose of our numerical experiment was to
check if an $\eta N$ repulsion could shift the peak to the right
and there is no information about such repulsion, we used the same
function $b(E)$ and did variations of $E_c$, namely $E_c/3$,
$E_c/2$, $E_c$, $2E_c$, and $3E_c$.
The corresponding curves are shown in Fig.~\ref{repuls.fig} where the 
larger $E_c$ the lower is the curve. All the curves depicted in 
Fig.~\ref{repuls.fig} were calculated with $a_{\eta N} = (0.75, 0.30)$ fm, 
$A = -0.85$ and $\kappa = \alpha = 3.316$ fm$^{-1}$.

 Therefore, comparison of our calculations with the
experimental data suggests that $A > -0.80$, Re $a_{\eta N} > 0.75$ fm, 
and the $\eta N$ interaction is likely to be repulsive at short
distances. Our calculations are highly sensitive to the
ratio of the photoproduction amplitudes for neutron and proton ($A$)
and depends rather weakly on the parameter $\kappa$.

\vspace{5mm}

Authors would like to thank Division for Scientific Affair of NATO for
support (grant CRG LG 970110) and DFG-RFBR for financial assistance
(grant 436 RUS 113/425/1).


\begin{thebibliography}{99}
\bibitem{Magi00} A. Magiera, H. Machner, Nucl. Phys {\bf A674} (2000) 515.

\bibitem{Hoeh98} G. Hoehler, $\pi$N Newsletters {\bf 14} (1998) 168.

\bibitem{Hai86} Q. Haider and L. C. Liu , Phys. Rev. Lett. {\bf B172} 
(1986) 257.

\bibitem{Rol96} M. R\"obig-Landau {\em et al.}, Phys. Lett. {\bf B373}
(1996) 45.

\bibitem{ref4} C. Wilkin, Phys. Lett. {\bf B331} (1994) 276.

\bibitem{ref6} H. C. Chiang, E. Oset, L. C. Liu, 
Phys. Rev. {\bf C44} (1991) 738.

\bibitem{ref7} G. L. Li, W. K. Cheng, T. T. S.  Kuo, 
Phys. Lett. {\bf B195} (1987) 515.

\bibitem{gnw} A. M. Green, J. A. Niskanen, S. Wycech, 
Phys. Rev. {\bf C54} (1996) 1970.

\bibitem{Ueda} T. Ueda, Phys. Rev. Lett. {\bf 66} (1991) 297.

\bibitem{Rak96} S. A. Rakityansky, S. A. Sofianos, M. Braun, 
V. B. Belyaev, and W. Sandhas, Phys. Rev. {\bf C53} (1996) R2043.

\bibitem{She98} N. V. Shevchenko, S. A. Rakityansky, S. A. Sofianos, 
V. B. Belyaev, and W. Sandhas,  Phys. Rev. {\bf C58} (1998) R3055.

\bibitem{Rit99} F. Ritz and H. Arenh\"ovel, 
Phys. Lett. {\bf B447} (1999) 15.

\bibitem{garcilazo} H. Garcilazo, Lett. Nuovo Cim. {\bf 28} (1980) 73.

\bibitem{bennh} C. Bennhold and H. Tanabe, Nucl. Phys. {\bf A530} 
(1991) 625.

\bibitem{PDG} Particle Data Group, Phys. Rev. {\bf D50} (1994) 1173.

\bibitem{batinic} M. Batinic, I. Slaus, A. Svarc, Phys. Rev. {\bf C52}, 
(1995) 2188.

\bibitem{finn} A. M. Green, S. Wycech, Phys.Rev. {\bf C55}
(1997) R2167.

\bibitem{speth} V. Yu. Grishina, L. A. Kondratyuk, M. Buescher, 
C. Hanhart, J. Haidenbauer, J. Speth,
Phys. Lett. {\bf B475} (2000) 9. 

\bibitem{Bel} V. B. Belyaev, {\it Lectures in Few-Body Systems}, 
Springer Verlag.

\bibitem{Soh71} F. Sohre and H. Ziegelman, Phys. Lett. {\bf B34} 
(1971) 579.

\bibitem{bib1} N. Hoshy, H. Hyuga and K. Kubodera, Nucl. Phys.
{\bf A324} (1979) 234.

\bibitem{bib2} D. Halderson and A. S. Rosenthal, Nucl. Phys. {\bf A501} 
(1989) 856.

\bibitem{bib3} Y. Zhang and D. Halderson, Phys. Rev. {\bf C45} (1992) 563.

\bibitem{bib4} E. Breitmoser, H. Arenhoevel, Nucl. Phys. {\bf A612} 
(1997) 321.

\bibitem{bib5} L. Tiator, C. Bennhold, S. S. Kamalov, Nucl. Phys. {\bf A580}
(1994) 455.

\bibitem{bib6} P. Hoffman-Rothe et al., Phys. Rev. Lett. {\bf 78} (1997) 4697.

\bibitem{Fix} A. Fix, H. Arenhoevel, Phys. Lett. {\bf B492} (2000) 32.

\bibitem{bib14} B. Krusche et al., Phys. Rev. Lett. {\bf 74} (1995) 3736.

\bibitem{bib15} A. M. Green and S. Wycech, Phys. Rev. {\bf C60} (1999) 035208.

\bibitem{bib16} N. C. Mukhopadhyay, J. F. Zhang, M. Benmerouche, 
Phys. Lett. {\bf B364} (1995) 1.

\end{thebibliography}
\end{document}